\documentclass[10pt,journal,twocolumn]{IEEEtran}
\IEEEoverridecommandlockouts
\usepackage{epsfig,cite,bm,algorithm,algorithmic,epstopdf,amsmath,subfigure,multirow,amssymb,amsfonts,amsthm,xcolor,caption,graphicx,textcomp,xcolor, subfigure}
\usepackage[inline, shortlabels]{enumitem}
\setlist{itemjoin ={,\enspace},itemjoin* = { and\enspace}}
\def\QEDclosed{\mbox{\rule[0pt]{1.5ex}{1.5ex}}}
\usepackage[T1]{fontenc}
\usepackage{stfloats}
\allowdisplaybreaks[4]

\allowdisplaybreaks[4]
\usepackage{booktabs}

\begin{document}

\title{Joint Uplink and Downlink Rate Splitting for Fog Computing-Enabled Internet of Medical Things}

\author{Jiasi Zhou, Yan Chen, Cong Zhou, Yanjing Sun,~\IEEEmembership{Member,~IEEE}, and Chintha Tellambura,~\IEEEmembership{Fellow,~IEEE}
\thanks{Jiasi Zhou is with the School of Medical Information and Engineering, Xuzhou Medical University, Xuzhou, 221004, China, (email: jiasi\_zhou@xzhmu.edu.cn). (\emph{Corresponding author: Jiasi Zhou}).}
\thanks{Yan Chen is with the Faculty of Information Technology and Electrical Engineering, University of Oulu, Oulu, Finland, (email: yan.chen@oulu.fi). }
\thanks{Cong Zhou is with the School of Electronic and Information Engineering, Harbin Institute of Technology, Harbin 150001, China, (email: zhoucong@stu.hit.edu.cn).}
\thanks{Yanjing Sun is with the School of Information and Control Engineering, China University of Mining and Technology, Xuzhou 221116, China (email: yjsun@cumt.edu.cn).}
\thanks{ Chintha Tellambura is with the Department of Electrical and Computer Engineering, University of Alberta, Edmonton, AB, T6G 2R3, Canada (email: ct4@ualberta.ca).} 
\thanks{This work was supported by the national key research and development program of China (2020YFC2006600) and the Talented Scientific Research Foundation of Xuzhou Medical University (RC20552228).}}
\maketitle
	
\begin{abstract}
The Internet of Medical Things (IoMT) facilitates in-home electronic healthcare, transforming traditional hospital-based medical examination approaches. This paper proposes a novel transmit scheme for fog computing-enabled IoMT that leverages uplink and downlink rate splitting (RS). Fog computing allows offloading partial computation tasks to the edge server and processing the remainder of the tasks locally. The uplink RS and downlink RS utilize their flexible interference management capabilities to suppress offloading and feedback delay. Our overarching goal is to minimize the total time cost for task offloading, data processing, and result feedback. The resulting problem requires the joint design of task offloading, computing resource allocation, uplink beamforming, downlink beamforming, and common rate allocation. To solve the formulated non-convex problem, we introduce several auxiliary variables and then construct accurate surrogates to smooth the achievable rate. Moreover, we derive the optimal computation resource allocation per user with closed-form expressions. On this basis, we recast the computing resource allocation and energy consumption at the base station to a convex constraint set. We finally develop an alternating optimization algorithm to update the auxiliary variable and inherent variable alternately. Simulation results show that our transmit scheme and algorithm exhibit considerable performance enhancements over several benchmarks.
\end{abstract} 

\begin{IEEEkeywords}
Internet of Medical Things, uplink rate splitting, downlink rate splitting, healthcare systems.
\end{IEEEkeywords}

\section{Introduction}

The Internet of Medical Things (IoMT) integrates medical devices and apps, collecting, exchanging, and utilizing healthcare data online. This enhances healthcare delivery, patient monitoring, and management, including fitness trackers and patient monitoring systems. IoMT utilizes biosensors for routine parameters like blood oxygen levels, blood pressure, etc., empowering personalized care, like glucose monitoring for diabetics. Continuous data analysis aids in early issue detection, minimizing hospital visits for chronic patients or the elderly. To revolutionize healthcare, IoMT relies on robust communications for real-time data and faces challenges in task offloading, data processing, and result feedback, which inspires our research \cite{9083675, 9490592, 9099795}.

Another driver of IoMT could be the rise of aging populations, which has led to a surge in interconnected medical devices, generating vast data. Traditional IoMT relies on cloud computing, involving data transfer to remote servers and processing before feedback to users \cite{9480807}, resulting in two drawbacks. Firstly, this method strains wireless links due to dual data transmission, compounded by extended communication distances causing transmit delays \cite{8730522}. Secondly, the heavy load of raw medical data can overwhelm cloud centers, causing processing delays and potentially risking outdated medical information \cite{8404033}. Consequently, cloud computing may not meet the demands of real-time and computation-heavy data loads.

Fog computing, as complementary to cloud computing, can provide pervasive low-latency computation services by decentralizing computing infrastructure\cite{9184073}. Data computing, storage, and applications are strategically distributed between the data source and the cloud, extending cloud services to the network edge where data originates, such as sensors or end-user devices. Its goal is to tackle issues like latency, bandwidth constraints, and real-time data analysis, which are crucial in IoMT applications with massive edge-generated data. By bringing computation and storage closer to data sources, fog computing reduces the need for long-distance data transmission to centralized centers, thus enhancing efficiency and response times. It complements cloud computing by focusing on real-time processing, low-latency interactions, and localized data storage. This allows for more efficient and responsive systems, especially in critical real-time decision-making scenarios.


 Fog computing-enabled medical devices are equipped with a certain amount of computation resources, which execute part of the medical analysis tasks at the source node\cite{qiu2021computation}. Therefore, it provides several attractive benefits, such as reducing the transmission of medical data, alleviating the burden on cloud centers, and improving the quality of experience. Recently, fog computing-enabled healthcare networks have been explored in \cite{9184073} and \cite{qiu2021computation}. However, due to limited computation sources and low-energy supply, individual medical devices may not process a large amount of collected data on short notice. Consequently, some of the computation tasks still need to be offloaded to the edge server or cloud server via wireless channels.

To reduce delays, individual devices must make optimal data offloading decisions and manage interference among multiple devices. The former requires optimal resource allocation algorithms, while the latter can be addressed through advanced multiple access schemes. These refer to the method or protocol that allows multiple devices or users to share the radio spectrum efficiently and concurrently. The most common ones are orthogonal multiple access (OMA), space division multiple access (SDMA)\cite{9083675,9309177}, or non-orthogonal multiple access (NOMA)\cite{askari2021energy,9181534}.  In OMA, the resource block (i.e., time, frequency, or code)  is divided into multiple non-overlapping slots, each allocated to a different device. This orthogonal setup completely averts co-channel interference\cite{10038476}.
SDMA allows all multiple users to transmit concurrently, which results in co-channel interference. Then, spatial multiplexing gains (e.g., beamforming with multiple antennas) can be utilized to counteract the interference. However, when the number of users exceeds the spatial degree of freedom (DoF) created through beamforming, co-channel interference contributes to a rate stagnation\cite{9835151}. NOMA is similar to SDMA on the transmit side but employs successive interference cancellation (SIC) on the receive side.  However, its performance gains strictly depend on complex receiver designs and large channel differences\cite{mao2018rate}.  When many users are scheduled per slot, the receiver complexity becomes intolerable for healthcare networks since medical devices have limited physical size. The aforementioned limitations call for a new approach. 

\subsection{Rate splitting (RS) approach}
RS,  a promising paradigm for sixth-generation (6G) networks, provides a more robust interference management solution.  This is achieved by enabling the receiver to decode partial interference while tolerating the residue, providing greater flexibility to counter interference\cite{10038476,mao2018rate,9831440}. RS can encapsulate  SDMA and NOMA as special cases by steering the decoding percentage. In general, uplink and downlink RS utilize different encoding and decoding strategies. Specifically, in uplink RS,   each message is divided into two sub-messages and then encoded into two separate streams. This process can be regarded as creating two virtual users\cite{10038476}. The receiver performs SIC to retrieve each stream and reconstruct the original message. In downlink RS,   each message is divided into common and private parts, all common parts are encoded into one common stream, and each private part is encoded into a private stream. Each receiver decodes the common stream, removes it via single-layer SIC, and then detects its desired private stream. Due to this flexible interference management, RS yields substantial performance gains under various user deployment scenarios\cite{10038476,mao2018rate,9831440}. Moreover, each user only needs a single-layer SIC. Therefore, RS is particularly suitable for low-complexity devices in healthcare networks.

Given their respective benefits, the synergy between fog computing and the RS scheme in IoMT may reap further gains and is a crucial topic. However, to our knowledge, applying the uplink and downlink RS schemes in fog computing-enabled IoMT remains unexplored. 

\subsection{Problem statement and contributions}\label{Section I.B}
To fill this gap, this paper proposes the joint uplink and downlink RS-based transmit scheme to alleviate the total time cost for medical data offloading, data processing, and result feedback in fog computing-enabled IoMT. We now elaborate on the time allocations of the different stages.    
\begin{enumerate}
	\item Each user decides whether to offload and how much computation task to offload to the BS, which has a server. The remaining portion will be processed locally by the user. After that, users use the uplink RS scheme to transmit their medical data to the BS. This stage takes time $T^u$.
	\item Once medical data are delivered, the server and users analyze the offloaded and remaining portions, respectively. This stage requires time $T^p$.
	\item After the server completes data processing, the BS uses the downlink RS scheme to feed back the processed result to users. This stage consumes time $T^d$.  
\end{enumerate}
These three stages must be executed in independent time slots since the $(n+1)$-th stage starts only when stage $n$ is over for $\forall n \in\{1,2\}$.   Thus, the total time cost is $T^u+T^p+T^d$. The scheduling process cannot be divided into three independent stages since the offloading decision goes through the overall process. This paper aims to minimize the total time cost, which is critical for healthcare applications.    However, this optimization goal presents several challenges. These include the determination of offloading decisions, the design of uplink and downlink beamforming, and the allocation of computation resources and common rates.  The resulting optimization problem is non-convex and NP-hard. Thus, it is not amenable to widely available convex algorithms. We propose a low-complexity,  iterative algorithm to address all these challenges, which at least yields a locally optimum solution. The main contributions are summarized as follows.
\begin{itemize}
\item We propose the joint uplink and downlink RS transmit scheme for fog computing-enabled IoMT. To minimize the total time cost $T^u+T^p+T^d$, we jointly optimize offloading policy, computation resource allocation, uplink beamforming, downlink beamforming, and common rate allocation. The minimization problem considers the constraints of the users' and BS's computational resources and energy consumption.
	
\item  The resulting optimization problem is non-convex and non-smooth, making it challenging to find a globally optimal solution using traditional convex methods. To tackle this issue, we adopt a surrogate optimization approach, which involves optimizing a simplified surrogate function instead of the original complex objective function. However, identifying accurate and easily optimizable surrogates presents its own set of challenges. To address this, we employ the quadratic transform method to create surrogates for uplink offloading and downlink feedback rates, introducing several auxiliary variables. Additionally, we determine the optimal allocation of computation resources at the user end, deriving a closed-form expression. Leveraging this, we transform the computation resource allocation and energy consumption at the BS  into a convex set. 

\item Subsequently, we frame the problem as a two-tier alternating optimization (AO) problem, where we iteratively optimize both the primary and auxiliary variables using convex optimization techniques and closed-form expressions. Throughout this process, we ensure convergence and optimality and consider the computational complexity of our proposed algorithm.
Our approach addresses the challenges by leveraging surrogate optimization, quadratic transformation, and convex optimization techniques to achieve efficient and effective resource allocations. We provide detailed discussions on convergence, optimality, and complexity to validate the efficacy of our proposed algorithm.
	
\item  We illustrate the superiority of our proposed RS-based transmit scheme and algorithm by extensive numerical simulation. The results show our scheme and algorithm yield substantial performance enhancements against several benchmarks and algorithms. 
\end{itemize}

\subsection{Related works}
During and after the COVID-19 pandemic, remote healthcare approaches have emerged to reduce infection risk and the healthcare burden.  This trend is facilitated by implementing IoMT, and several efforts have focused on the design of transmit schemes and resource allocation strategies \cite{9801732,ahmed2022physical,askari2021energy,9181534}. For example, federated learning-based algorithms can address the privacy and security issues of medical data\cite{9801732,ahmed2022physical}. The authors in \cite{askari2021energy} divide IoMT into two sub-networks: intra-wireless body area network (WBAN) and beyond-WBAN. Then, they utilize NOMA to boost connectivity levels and propose a resource scheduling algorithm. The NOMA-enabled IoMT is extended to multi-carrier scenarios in \cite{9181534}, where sparse vector coding is considered to suppress inter-carrier interference. However, employing the above transmit schemes, the collected medical information must be transmitted to the cloud computing center for processing\cite{9801732,ahmed2022physical,askari2021energy,9181534}.

To ease cloud centers' burden, many works consider edge computing- or fog computing-enabled IoMT\cite{9083675,9309177,9380652,9416297,9186655,9184073}. Similar to \cite{askari2021energy}, references \cite{9083675} and \cite{9309177} separate IoMT into intra-WBAN and beyond-WBAN, and then model these two sub-networks as a cooperative game and a non-cooperative game, respectively. In fog computing-enabled IoMT, optimized offloading decision helps boost network performance. Reference \cite{9380652} divides the task offloading and resource allocation problem into three sub-problems and then proposes the corresponding algorithms to attack them. The authors in \cite{9416297} formulate the task offloading policy as a Stackelberg game and propose an iterative algorithm to make the smartest offloading decision. Besides, one can model the task offloading as an adversarial multi-armed bandit problem and use machine learning to derive the long-term optimal offloading strategy\cite{9186655}. Considering network connection uncertainty incurred by user mobility, one can formulate the offloading problem as multi-stage stochastic programming and employ the sample average approximation approach to obtain sub-optimal offloading decisions\cite{9184073}. These efforts identify the optimal or sub-optimal offloading decision but utilize SDMA to offload medical data. This brings the saturated transmit rate and thus compromises the response time. 

RS scheme conquers the rate saturation phenomenon through flexible interference management. Many works have demonstrated that downlink RS presents considerable benefits over the hitherto multiple access schemes under various user deployments, such as overloaded deployment\cite{9382277}. These benefits include higher energy efficiency\cite{10032267}, higher fairness rate\cite{9991090}, better robustness\cite{9786780}, and higher DoF\cite{7434643}. Given its benefits, the RS scheme has been extended to many communication systems/networks, such as unmanned aerial vehicles (UAV) networks\cite{10269133}, visible light communication (VLC) networks\cite{9693949}, cloud radio access networks (C-RAN)\cite{9445019}, multibeam satellite systems\cite{9790069}, and intelligent reflecting surface (IRS) systems\cite{9912342}. The above contributions consider the downlink RS scheme, but several recent works discover that the uplink RS scheme improves capacity regions without requiring time sharing among users\cite{10038476}.  For example, uplink RS exhibits lower outage probability than NOMA in two-user systems\cite{9755045,9676684}, short packet communication systems\cite{xu2022rate}, or cognitive radio systems\cite{yue2022ergodic}. To further expose its benefits, reference \cite{yang2020sum} extends uplink RS to multi-user systems and maximizes the sum rate. With the same objective, references \cite{9691467} and \cite{10014691} consider uplink RS-aided multi-user multi-antenna systems. In addition to boosting performance, uplink RS can also reduce electromagnetic exposure risk\cite{10032157}. Recently, uplink RS has been considered in multi-antenna healthcare systems\cite{10285055,10251435}. These two works utilize the uplink RS scheme to prevent medical information leakage and transfer medical data. However, all the collected medical data must be offloaded to the BS in\cite{10251435} while the feedback of processing result has not been considered. Integrating uplink and downlink RS for fog computing-enabled communication systems has been missing in existing works.

\emph{Organization:}  The remainder of this paper is organized as follows.  Section \ref{Section II} elaborates on the system model and formulates the optimization problem. Section \ref{Section III} presents the proposed iterative optimization algorithm and its properties analysis.  Section \ref{Section IV} provides simulation results. Section \ref{Section V} concludes the whole paper. 

\emph{Notations:} Let boldface upper-case letters, boldface lower-case letters, and calligraphy letters denote matrices, vectors, and sets. ${(\bullet)}^H$ represents the Hermitian transpose. $\mathrm{Re}\left( \bullet\right)$ and $\mathrm{Tr}\left( \bullet\right)$ denote the real part and trace. $\mathcal{CN}(\mu, \sigma^2)$ denotes a complex Gaussian of mean $\mu$ and variance $\sigma^2$.

\section{System model and problem formulation}\label{Section II}
\begin{figure}[tbp]
\centering
\includegraphics[scale=1.2]{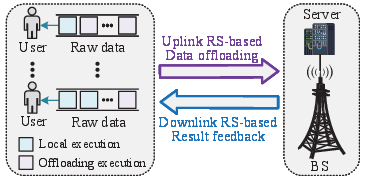}
\caption{An uplink and downlink RS-based transmit model for fog computing-enabled IoMT.}
\label{fig:system}
\end{figure}
As mentioned, we consider a fog computing-enabled IoMT -- Fig.~\ref{fig:system}, which comprises $K$ users with limited computation capacity and one BS with an edge server. Let $\mathcal{K}=\{1,\dots,K\}$ denote users index set. Each user has $A_{ut}>1$ transmit antennas and a single receive antenna, while the BS has $A_{bt}>1$ transmit antennas and $A_{br}>1$ receive antennas. Denote the channel matrix from the $k$-th user to the BS as $\mathbf{H}_k\in\mathbb C^{A_{ut}\times A_{br}}$. Similarly, the channel vector from the BS to the $k$-th user is denoted by $\mathbf{h}_k\in\mathbb C^{A_{bt}\times 1}$. This paper assumes channel state information (CSI) is perfectly available to the BS and users. A similar assumption has been widely made in \cite{10251435,9309177,9184073}. Due to the limited computation capacity, these users aim to offload their collected physiological data to the server for medical analysis and get feedback rapidly. As a result, the overall medical process has three scheduling phases: task offloading, data processing, and result feedback, which has been summarized as Section \ref{Section I.B}. Based on the optimized offloading decision, partial or full computing tasks would be offloaded to the server. Our model embraces partial and full offloading as two special cases.   We propose the uplink and downlink RS transmit schemes to ease task offloading and feedback time. Next, we will outline these three data transmission or processing models.

\subsection{Rate splitting-based uplink transmission model}
To offload the computing task in time, $k$-th user can transmit a $A_u$-dimensional vector of messages $\mathbf{s}_k=\left[S^{(1)}_k,\dots, S^{(A_u)}_k\right]$ to the BS, where $A_u\leq\min{\left(A_{ut},A_{br}\right)}$. Following the uplink RS scheme, at $k$-th user, each of its $A_u$ messages $S^{(a)}_k$ for $a\in\{1,\dots,A_u\}$ is split into two parts, $S^{(a)}_{k,m}$ for $\forall m\in\{1,2\}$, which is encoded independently into $x^{(a)}_{k,m}$\cite{10038476}. Then, the $A_u$-dimensional vector $\mathbf{x}_{k,m}=\left[x^{(1)}_{k,m},\dots,x^{(A_u)}_{k,m}\right]^T$ is linearly precoded by $\mathbf{W}_{k,m} \in\mathbb C^{A_{ut}\times A_u}$. These two streams are superposed to form $\mathbf{x}_k$, so the transmit signal at $k$-th user is $\mathbf{x}_k=\sum_{m=1}^{2}\mathbf{W}_{k,m} \mathbf{x}_{k,m}$. As such, the signal received at the BS is 
\begin{equation}
\mathbf{y}=\sum_{k=1}^{K}\sum_{m=1}^{2}\mathbf{H}^H_{k}\mathbf{W}_{k,m} \mathbf{x}_{k,m} + \mathbf{n},
\end{equation}
where $\mathbf{n}\sim \mathcal{CN}\left(\mathbf{0},\sigma^2\mathbf{I}_{A_{br}}\right)$ is the additive white Gaussian noise (AWGN) vector. Then, the BS utilizes SIC to retrieve and reconstruct the original signal based on a predetermined decoding order. Consequently, when the BS decodes $\mathbf{x}_{k,m}$, the interference power is
\begin{align}
\bm{\Omega}_{k,m} =&\sum\nolimits_{\bm{\pi}_{k,m}}\mathbf{H}^H_{i}\mathbf{W}_{i,j}\mathbf{W}^H_{i,j}\mathbf{H}_{i}+ \mathbf{I}_{A_{br}},
\label{Interference}
\end{align}
where set $\bm{\pi}_{k,m}=\{i,j|i\in \mathcal{K}, j\in\{1,2\},\pi_{i,j}>\pi_{k,m}\}$ denotes the indexes of the stream $\mathbf{x}_{i,j}$ is decoded after stream $\mathbf{x}_{k,m}$ and $\pi_{k,m}$ is the decoding order of stream $\mathbf{x}_{k,m}$. Our previous work \cite{10251435} develops an iterative algorithm to optimize the decoding order, which yields the optimal ordering. Readers are referred to \cite{10251435} for more details about ordering optimization. This paper optimizes the offloading policy, uplink beamforming, downlink beamforming, and computation resource allocation to minimize the total time cost.  In \cite{10251435}, we only study the uplink RS scheme to handle the data communication needs. In contrast, this work makes two advancements. First, we introduce fog computing techniques to ease the burden on the server and thus shorten the transmission distance of medical data. Secondly, we consider the uplink task offloading and the downlink feedback of the processed result, complicating the co-channel interference management. According to the decoding principle of the uplink RS scheme, the signal-to-interference-plus-noise
ratio (SINR) can be written as
\begin{equation}
\bm{\Gamma}_{k,m}=\mathbf{W}^{H}_{k,m}\mathbf{H}_{k}\left(\bm{\Omega}_{k,m}\right)^{-1}\mathbf{H}^H_{k}\mathbf{W}_{k,m}.
\label{SINR}
\end{equation}

Using Gaussian codebooks, the achievable rate of decoding $\mathbf{x}_{k,m}$ is
\begin{equation}
R^u_{k,m}=\log\det\left(\mathbf{I}+\bm{\Gamma}_{k,m}\right).
\label{Rate}
\end{equation}
Thus, the total task offloading rate is $R^u_k=R^u_{k,1}+R^u_{k,2}$. 

Similar to \cite{8638800} and\cite{9186655}, we employ a task-partition model, where the computation task can be further partitioned into several subtasks. This assumption is reasonable since a computing task generally comprises smaller subtasks in IoMT. For example, the data collected by users comprises blood oxygen, blood pressure, temperature, and so on.  Let $L_k$ and $\alpha_k$ denote the collected medical data size and the offloading percentage at $k$-th user, where $\alpha_k\in[0,1]$.
 As a result, the offloading delay of $k$-th user can be expressed as
\begin{equation}
T^u_{k} = \frac{\alpha_k L_k}{R^u_k}.
\label{Rate}
\end{equation}
When all users have finished task offloading, the offloading stage ends and the data processing stage starts. As such, the offloading stage depends on the maximum among all users, so the task offloading delay is $T^u=\max_{\forall k\in\mathcal{K}}T^u_k$.

\subsection{Data processing model}
After receiving the physiological data packets from the users, the BS begins to process and compute them. The computing time cost for $k$-th user is
\begin{equation}
T^s_{k} = \frac{\omega_k\alpha_k L_k}{f_k},
\end{equation}
where $\omega_k$ is the number of CPU cycles required for processing one byte of task of $k$-th user (abbreviated as task $k$). $f_k$ is the CPU-cycle frequency assigned for processing the task $k$. The computing energy consumption for executing task $k$ is \cite{9416297}
\begin{equation}
E^s_{k} = \kappa f^2_k\omega_k\alpha_k L_k,
\end{equation}
where $\kappa$ is the energy coefficient, dependent on the chip architecture. Meanwhile, each user executes the remaining task $(1-\alpha)L_k$ locally. The corresponding time cost is
\begin{equation}
T^l_{k} = \frac{\left(1-\alpha_k\right)\omega_k L_k}{\tilde f_k},
\label{Time_l}
\end{equation}
where $\tilde f_k$ is the CPU-cycle frequency of local computing at the $k$-th user. Similarly, the energy consumption is 
\begin{equation}
E^l_{k} = \kappa \tilde f^2_k\omega_k(1-\alpha_k) L_k.
\label{Energy_l}
\end{equation}
The BS can feed the result to users until all the computation tasks are handled. Thus, the data processing time depends on the maximum among $T^s_k$ and $T^l_k$ for $\forall k$, which is 
\begin{equation}
T^p = \max_{\forall k\in\mathcal{K}}\left[\max\left(T^s_k, T^l_k\right)\right].
\label{Time:data processing}
\end{equation}

\subsection{Rate splitting-based downlink transmission model}
After finishing task computation, the BS uses the downlink RS scheme to feed the results to the users. With the downlink RS approach, message $W_k$ for user $k$ is split into a common part $W_{k,c}$ and a private part $W_{k,p}$ for $\forall k$. All the common parts are combined and encoded to one common stream $x_c$ while private parts are encoded into $K$ private stream $x_k$ for $\forall k$. These streams are mutually independent and linearly precoded by $\mathbf{p}_c\in\mathbb C^{A_{bt}\times 1}$ and $\mathbf{p}_k\in\mathbb C^{A_{bt}\times 1}$. Then, the common stream is superposed on top of all private streams in a non-orthogonal manner. Thus, the transmitted signal is $\mathbf{x}=\mathbf{p}_cx_c + \sum_{k=1}^{K}\mathbf{p}_{k}x_k$. Consequently, the received signal at the $k$-th user can be expressed as
\begin{equation}
y_k=\mathbf{h}^H_{k}\mathbf{p}_cx_c + \mathbf{h}^H_{k}\sum_{k=1}^{K}\mathbf{p}_{k}x_k +n_k,
\end{equation}
where $n_k\sim \mathcal{CN}\left(0,\sigma^2_k\right)$ denotes AWGN term. The average power received by the $k$-th user is 
\begin{equation}
T_{k,c}=\overbrace{{\left|\mathbf{h}^H_{k}\mathbf{p}_c\right|}^2}^{S_{k,c}}+\underbrace{\overbrace{{\left|\mathbf{h}^H_{k}\mathbf{p}_{k}\right|}^2}^{S_{{k,p}}}+\overbrace{\sum_{j=1,j\neq k}^{K}\left|\mathbf{h}^H_{k}\mathbf{p}_j\right|^2+\sigma^2_k}^{I_{{k,p}}}}_{I_{k,c}=T_{k,p}}.\label{equ:power}
\end{equation}

To recover the desired message, the $k$-th user with 1-layer SIC  decodes the common stream by treating all private streams as noise and then removes it via SIC, so the SINR is $\gamma_{k,c}=S_{k,c}{I^{-1}_{k,c}}$. However, to ensure that all users can successfully recover $x_c$, the common rate shall not exceed $R^d_c= \min_{\forall k}\log\left(1+\gamma_{k,c}\right)$. Furthermore, the common rate is shared by all users, so we have $R^d_c=\sum_{k=1}^{K}R^d_{k,c}$, where $R^d_{k,c}$ is the portion of common rate allocated to $W_{k,c}$. After removing the common stream via SIC, the $k$-th user decodes its desired private stream by directly treating other private streams as noise. The resultant SINR and decoding rate are $\gamma_{{k,p}}=S_{k,p}{I^{-1}_{{k,p}}}$ and $R^d_{k,p}=\log\left(1 + \gamma_{{k,p}}\right)$, respectively. The overall feedback rate of $k$-th user is
$R^d_k=R^d_{k,c}+R^d_{k,p}$. Therefore, the feedback time of $k$-th user is 
\begin{equation}
T^d_{k} = \frac{\epsilon\alpha_k L_k}{R^d_k},
\label{Time}
\end{equation}
where $\epsilon$ denotes the compression parameter. When all the users receive the feedback data, this stage terminates, so the feedback time cost is $T^d=\max_{\forall k} T^d_{k}$.

\subsection{Problem formulation}
This paper aims to minimize total time cost by jointly designing offloading policy ${\bm{\alpha}} = \left\{\alpha_k\right\}$, computation resource allocation $\mathbf{f} =  \left\{f_k\right\}$ and $\tilde{\mathbf{f}} = \left\{\tilde f_k\right\}$, uplink beamforming $\mathbf{W}=\left\{\mathbf{W}_{k,m}\right\}$, downlink
beamforming $\mathbf{p}=\left\{\mathbf{p}_c,\mathbf{p}_k\right\}$, and common rate allocation $\mathbf{r} = \left\{R^d_{k,c}\right\}$. This problem is formulated as
\begin{subequations}\label{linear_p}
	\begin{align}
&\min_{\mathbf{W},\mathbf{p},\mathbf{f},\tilde{\mathbf{f}},\mathbf{r},{\bm{\alpha}}}T^u+T^p+T^d,\label{ob_a}\\
	\text{s.t.}\quad
	&\sum_{m=1}^{2}||\mathbf{W}_{k,m}||^2\leq P_k,~\forall k,\label{ob_b}\\
 &||\mathbf{p}_{c}||^2 + \sum_{k=1}^{K}||\mathbf{p}_{k}||^2\leq P_b,\label{ob_c}\\
        &\sum_{k=1}^{K}f_k\leq F_b,\label{ob_d}\\
        &\sum_{k=1}^{K}E^s_k\leq P_bT^p,\label{ob_e}\\
        &\tilde f_k\leq F_k,~\forall k,\label{ob_f}\\
        &E^l_k\leq P_kT^p,~\forall k,\label{ob_g}\\
        &R^d_{k,c}\geq 0,~\forall k,\label{ob_h}\\
        &\alpha_k\in[0,1],~\forall k,\label{ob_i}
	\end{align}
\end{subequations}
where $P_k$ and $F_k$ are the transmit power and computation capacity thresholds at $k$-th user, respectively. Similarly, $P_b$ and $F_b$ are the corresponding thresholds at the BS. (\ref{ob_b}) and (\ref{ob_c}) are the transmit power constraints at $k$-th user and the BS, respectively. (\ref{ob_d}) and (\ref{ob_e}) are the transmit power and computation capacity constraints at the BS end. (\ref{ob_f}) and (\ref{ob_g}) are the corresponding constraints per user. Note that the transmit power at the BS is $P_b$, so the available energy is $P_bT^p$ during the data processing period. Similarly, the energy provided by $k$-th user is  $P_kT^p$. (\ref{ob_h}) ensures the common rate allocation parameter is non-negative. (\ref{ob_i}) is the offloading percentage constraint.

Problem (\ref{linear_p}) exhibits high non-convexity and non-smoothness, posing three technical challenges. Firstly, time cost in the three scheduling phases and transmit rate are non-convex. Such problems are difficult to solve in the primal and dual domains because of the unknown duality gap. Secondly, the maximum operators incur non-smoothness, especially data processing time $T^p$, complicating the optimization process. Thirdly, the intricate coupling between beamforming and computation resource allocation variables aggravates the difficulty. Consequently, the global optimal solution is hopelessly elusive.

\section{Proposed algorithm and properties analysis}\label{Section III}
This section uses \textbf{surrogate optimization} to attack the non-convex problem (\ref{linear_p}). When evaluating the objective function can be time-consuming or costly, we can construct a surrogate model, a simplified representation of the original function. This surrogate model is then used to approximate the behavior of the objective function, allowing for faster and more efficient optimization. Its technical challenge is how to create accurate and easily optimized surrogates. 
To construct surrogates for the offloading and feedback rates, we employ the quadratic transform method and introduce auxiliary variables. Moreover, we transform the computation resource and energy consumption constraints (\ref{ob_d}), (\ref{ob_e}), (\ref{ob_f}), and (\ref{ob_g}) to a tractable form via first-order Taylor expansion. We then treat the recast problem as a two-tier AO problem, where variables are partitioned into two blocks, namely, inherent variables $\mathcal{Q}_1 = \{\mathbf{W},\mathbf{P},\mathbf{f},\tilde{\mathbf{f}}\}$ and auxiliary variables $\mathcal{Q}_2 = \{\mathbf{y},{\mathbf{Y}},{\mathbf{\Phi}}\}$.  The convex framework and closed-form expressions solve the inherent and auxiliary variable blocks, respectively. We finally discuss several crucial properties of our proposed algorithm, including convergence, optimality, and complexity.

In the problem (\ref{linear_p}), fractional SINR and coupled beamforming in logarithmic functions hamper optimization. The classic approach to solving similar problems is the semi-definite relaxation (SDR)-based iterative algorithm. This method decouples the interleaved beamforming by introducing a rank-one semi-definite matrix and then overlooking the rank-one constraint when solving the recast problem. The SDR-based algorithm exhibits good performance but has two drawbacks. Firstly, its complexity is $\mathcal{O}\left(KA^7_{t}\right) $, where $K$ and $A_t$ denote the number of users and beamforming dimension, respectively. Thus, its computational cost is too high, especially in multi-user scheduling scenarios. Secondly, the solution may not be feasible for the optimization problem since the rank-one constraint is omitted. These limitations call for a less complex algorithm. To this end, we use the quadratic transform approach to decouple the fractional SINR into a difference of two terms first proposed in \cite{shen2018fractional}. Theorem 2 in \cite{shen2018fractional} motivates our Claim 1.

{\bf\emph{Claim 1}}: By introducing  our constructed surrogte function $f\left(\mathbf{y},\mathbf{p}\right) = \log\left(1+ 2\mathrm{Re}\left(\mathbf{y}^{H}\mathbf{s}\left(\mathbf{p}\right)\right)-\mathbf{y}^{H}I\left(\mathbf{p}\right)\mathbf{y}\right)$ for any $\mathbf{s}(\mathbf{p})\in\mathbb C^{A_{bt}\times 1}$ and $I(\mathbf{p}) > 0$, one has 
\begin{equation}
\log\left(1+\frac{(\mathbf{s}\left(\mathbf{p}\right))^H\mathbf{s}\left(\mathbf{p}\right)}{I\left(\mathbf{p}\right)}\right)=\max_{\mathbf{y}}f\left(\mathbf{y},\mathbf{p}\right) 
\label{Surrogate}
\end{equation} 
and the optimal solution to the right-hand of equation (\ref{Surrogate}) is $\mathbf{y}^*=\left(I\left(\mathbf{p}\right)\right)^{-1}\mathbf{s}\left(\mathbf{p}\right)$.

\emph{Proof}: We observe that $f\left(\mathbf{y},\mathbf{p}\right)$ is a concave function over $\mathbf{y}$ since $I\left(\mathbf{p}\right)>0$ and the logarithm function is increasing. As such, we can derive optimal $\mathbf{y}^*$ via $\frac{\partial f\left(\mathbf{y},\mathbf{p}\right)}{\partial \mathbf{y}}=0$. We thus get $\mathbf{y}^*=\left(I\left(\mathbf{p}\right)\right)^{-1}\mathbf{s}\left(\mathbf{p}\right)$. Plugging $\mathbf{y}^*$ into $f\left(\mathbf{y},\mathbf{p}\right)$, we have that the left-hand of equation (\ref{Surrogate}) and its surrogate are strictly equivalent. \hfill \QEDclosed

Engaging Claim 1 to recast one-dimensional logarithmic functions $R^d_{k,c}$ and $R^d_{k,p}$, we construct two surrogates, as follows.
\begin{subequations}\label{FP_1}
\begin{align}
&f_{k,c}\left(\mathbf{y}_{k,c},\mathbf{p}\right)= \log\Big(1+2\mathrm{Re}\left(\mathbf{y}^H_{k,c}\mathbf{h}^H_{k}\mathbf{p}_{c}\right)-\mathbf{y}^H_{k,c}I_{k,c}\mathbf{y}_{k,c}\Big),\label{f_c2_mk}\\
&f_{k,p}\left(\mathbf{y}_{k,p},\mathbf{p}\right)= \log\Big(1+2\mathrm{Re}\left(\mathbf{y}^H_{k,p}\mathbf{h}^H_{k}\mathbf{p}_{k}\right)-\mathbf{y}^H_{k,p}I_{k,p}\mathbf{y}_{k,p}\Big).\label{f_c1_mk}
\end{align}
\end{subequations}

Our surrogate function does not seem to define a convex set, but it will be clear later that it can be easily solved by an alternating method. Claim 1 can create an accurate surrogate for one-dimensional beamforming vectors, but it is unsuitable for multi-dimensional beamforming matrices. Multi-dimensional matrix optimization is more complex than one-dimensional vector optimization. We develop Claim 2, driven by \cite{8862850} to construct surrogates for the offloading rate.

\begin{table*}
	\hrule
	\begin{subequations}\label{FP_4}
	\begin{align}
&\tilde f\left(\mathbf{Y},\mathbf{\Phi},\mathbf{W}\right) = \log\det\left(\mathbf{I}+\bm{\Phi}\right)+\frac{1}{\ln 2}\mathrm{Tr}\left(\left(\mathbf{I}+\bm{\Phi}\right)\left(2\mathbf{S}(\mathbf{W})\mathbf{Y}-\mathbf{Y}^H\left(\mathbf{S}^H(\mathbf{W})\mathbf{S}(\mathbf{W})+\bm{\Omega}(\mathbf{W})\right)\mathbf{Y}\right)-\bm{\Phi}\right);\label{Surrogate_3}\\
&\tilde f_1\left(\mathbf{\Phi},\mathbf{W}\right) = \log\det\left(\mathbf{I}+\bm{\Phi}\right)+\frac{1}{\ln 2}\mathrm{Tr}\left(\left(\mathbf{I}+\bm{\Phi}\right)\left(\mathbf{S}(\mathbf{W})\left(\bm{\Omega}(\mathbf{W})+\mathbf{S}^H(\mathbf{W})\mathbf{S}(\mathbf{W})\right)^{-1}\mathbf{S}^H(\mathbf{W})\right)-\bm{\Phi}\right);\label{Surrogate_4}\\
&\tilde f_2\left(\mathbf{\Phi},\mathbf{W}\right) = \log\det\left(\mathbf{I}+\bm{\Phi}\right)-\frac{1}{\ln 2}\mathrm{Tr}\left(\left(\mathbf{I}+\bm{\Phi}\right)\left(\mathbf{I}+\mathbf{S}(\mathbf{W})\left(\mathbf{\Omega}(\mathbf{W})\right)^{-1}(\mathbf{W})\mathbf{S}^H(\mathbf{W})\right)^{-1}\right) + \frac{1}{\ln 2}A_{r}.\label{Surrogate_5}
	\end{align}
\end{subequations}
\hrule
\end{table*}

{\bf\emph{Claim 2}}: By introducing  our constructed surrogate function $\tilde f\left(\mathbf{Y},\mathbf{\Phi},\mathbf{W}\right)$ for any $\mathbf{S}(\mathbf{W})\in\mathbb C^{A_t\times A_r}$ and $\mathbf{\Omega}(\mathbf{W}) \succ\mathbf{0}$, where $\tilde f\left(\mathbf{Y},\mathbf{\Phi},\mathbf{W}\right)$ is as (\ref{Surrogate_3}) shown at the top of this page, one has
\begin{equation}
\log\det\left(\mathbf{I}+\frac{\mathbf{S}^H(\mathbf{W})\mathbf{S}(\mathbf{W})}{\mathbf{\Omega}(\mathbf{W}) }\right) = \max_{\mathbf{Y},\mathbf{\Phi}}\tilde f\left(\mathbf{Y},\mathbf{\Phi},\mathbf{W}\right)
\label{equ:Surrogate_2}
\end{equation} 
and the optimal solution to the right-hand of equation (\ref{equ:Surrogate_2}) is $\mathbf{Y}^*=\left(\mathbf{S}^H(\mathbf{W})\mathbf{S}(\mathbf{W}) +\mathbf{\Omega}(\mathbf{W}) \right)^{-1}\mathbf{S}(\mathbf{W})$ and $\mathbf{\Phi}^*=\mathbf{S}(\mathbf{W})\mathbf{\Omega}(\mathbf{W})^{-1}\mathbf{S}^H(\mathbf{W})$.

\emph{Proof}: We first observe that $\mathrm{Tr}\left(\left(\mathbf{I}+\bm{\Phi}\right)\mathbf{X}\right)$ is an increasing function over $\mathbf{X}$, so $\mathrm{Tr}\left(\left(\mathbf{I}+\bm{\Phi}\right)\mathbf{X}\right)$ is concave over $\mathbf{Y}$, where $\mathbf{X}=2\mathbf{S}(\mathbf{W})\mathbf{Y}-\mathbf{Y}^H\left(\mathbf{S}^H(\mathbf{W})\mathbf{S}(\mathbf{W})+\bm{\Omega}\right)\mathbf{Y}$. Similar to Claim 1, $\tilde f\left(\mathbf{Y},\mathbf{\Phi},\mathbf{W}\right)$ can be maximized when 
\begin{equation}
\mathbf{Y}^* = \left(\mathbf{S}^H(\mathbf{W})\mathbf{S}(\mathbf{W})+\bm{\Omega}(\mathbf{W})\right)^{-1}\mathbf{S}(\mathbf{W}). 
\end{equation}
Plugging $\mathbf{Y}^*$ into (\ref{Surrogate_3}), we get (\ref{Surrogate_4}). According to the Woodbury matrix identity, we rewrite (\ref{Surrogate_4}) as (\ref{Surrogate_5}), which is concave function over $\mathbf{\Phi}$. As a result, $\frac{\partial \tilde f_2\left(\mathbf{\Phi},\mathbf{W}\right)}{\partial \mathbf{\Phi}}=0$ ensures that $\tilde f_2\left(\mathbf{\Phi},\mathbf{W}\right)$ can be maximized, so we have
\begin{equation}
\left(\mathbf{I}+\mathbf{S}(\mathbf{W})\left(\mathbf{\Omega}(\mathbf{W}) \right)^{-1}\mathbf{S}^H(\mathbf{W})\right)^{-1} = \left(\mathbf{I}+ \mathbf{\Phi}\right)^{-1}.
\end{equation} 
We thus derive $\mathbf{\Phi}^*=\mathbf{S}(\mathbf{W})\left(\mathbf{\Omega}(\mathbf{W})\right)^{-1}\mathbf{S}^H(\mathbf{W})$. Plugging $\mathbf{\Phi}^*$ into $\tilde f_2\left(\mathbf{\Phi},\mathbf{W}\right)$, Claim 2 can be proved.\hfill \QEDclosed

We employ Claim 2 to construct a surrogate for the offloading rate $R^u_{k,m}$, which is defined as (\ref{Surrogate_6}) shown at the top of the next page. In addition to the fractional SINR, the fractional time cost is another difficulty in solving the problem (\ref{linear_p}). Fortunately, a ratio of quadratic over linear polynomial is convex, which is amenable for easy solutions. Therefore, we replace the offloading percentage $\alpha_k$ with a quadratic variable say $\beta^2_k$, where $\beta_k\in\left[0,1\right]$. Note that these two variable representation methods are correct and completely equivalent. However, the fractional time cost is hard to solve using the former approach. After that, to attack the non-smoothness, we remove the maximum operators in the problem (\ref{linear_p}). The recast problem is as follows.
\begin{table*}	
\hrule
\begin{align}
\tilde f\left(\mathbf{Y}_{k,m},\mathbf{\Phi}_{k,m},\mathbf{W}\right) = \log\det\left(\mathbf{I}+\bm{\Phi}_{k,m}\right)+\frac{1}{\ln 2}\mathrm{Tr}\left(\left(\mathbf{I}+\bm{\Phi}_{k,m}\right)\left(2\mathbf{W}^H_{k,m}\mathbf{H}_k\mathbf{Y}_{k,m}-\mathbf{Y}^H_{k,m}\left(\mathbf{H}^H_k\mathbf{W}_{k,m}\mathbf{W}^H_{k,m}\mathbf{H}_k+\bm{\Omega}_{k,m}\right)\mathbf{Y}_{k,m}\right)-\bm{\Phi}_{k,m}\right).\label{Surrogate_6}
\end{align}
\hrule
\end{table*}
\begin{subequations}\label{linear_p2}
	\begin{align}
&\min_{\mathbf{W},\mathbf{p},\mathbf{f},\tilde{\mathbf{f}},\mathbf{R},{\bm{\beta}},\mathbf{T}}T^u+T^p+T^d,\label{ob_a2}\\
	\text{s.t.}\quad
	&T^u\geq \frac{\beta^2_k L_k}{R^u_{k,1}+R^u_{k,2}},~\forall k,\label{ob_b2}\\
 &T^p\geq \frac{\omega_k\beta^2_kL_k}{f_k},~\forall k,\label{ob_c2}\\
        &T^p\geq \frac{\omega_k(1-\beta^2_k)L_k}{\tilde f_k},~\forall k,\label{ob_d2}\\
        &T^d\geq \frac{\epsilon\beta^2_kL_k}{R^d_{k,c}+R^d_{k,p}},~\forall k,\label{ob_e2}\\
        &\sum_{i=1}^{K}R^d_{i,c}\leq \max_{\mathbf{y}_{k,c}}f_{k,c}\left(\mathbf{y}_{k,c},\mathbf{p}\right),~\forall k\label{ob_f2}\\
        &R^d_{k,p}\leq \max_{\mathbf{y}_{k,p}} f_{k,p}\left(\mathbf{y}_{k,p},\mathbf{p}\right),~\forall k,\label{ob_g2}\\
        &R^u_{k,m}\leq \max_{\mathbf{Y}_{k,m},\mathbf{\Phi}_{k,m}} \tilde f\left(\mathbf{Y}_{k,m},\mathbf{\Phi}_{k,m},\mathbf{W}\right),~\forall k,m,\label{ob_h2}\\
        &\beta_k\in\left[0,1\right],\forall k,\label{ob_i2}\\
        &\mbox{(\ref{ob_b}),~(\ref{ob_c}),~(\ref{ob_d}),~(\ref{ob_e}),~(\ref{ob_f}),~(\ref{ob_g}),~(\ref{ob_h}),}\label{ob_j2}
	\end{align}
\end{subequations}
where ${\bm{\beta}} = \{\beta_k\}$ and $\mathbf{R}=\left\{R^d_{k,p},R^d_{k,c},R^u_{k,m}\right\}$. The problem (\ref{linear_p2}) involves the mutual coupling between the offloading percentage and computation resource allocation variables in (\ref{ob_d2}) and (\ref{ob_i2}), which still appears intractable. This drives us to develop Claim 3, which helps recast constraints (\ref{ob_d2}) and (\ref{ob_i2}).

{\bf\emph{Claim 3}}: We can derive two conclusions in the data processing phase. Firstly, the time cost $T^p$ depends on the local computing time $T^l_k$. Secondly, the local computing time cost among all users is strictly equal. These two conclusions mean $T^l_k=T^p$ for $\forall k$. 

\emph{Proof}: We now prove Claim 3 by contradiction. We assume that computing time locally does not decide the data processing time cost, \emph{i.e.}, $T^p=\max_{\forall k}T^s_k$. Without loss of generality, let $T^{s}_{k}$ denote the maximum time cost. Meanwhile, the local computing time cost and resource allocation at the $k$-th user is denoted by $T^l_k$ and $\tilde f_k$, respectively. Since the resource allocation can meet constraints (\ref{ob_f}), (\ref{ob_g}), and equation (\ref{Time:data processing}), we have 
 \begin{subequations}\label{contradict}
	\begin{align}
	 &E^l_{k} = \kappa \tilde f^2_k\omega_k(1-\alpha_k) L_k\leq P_kT^p,\label{Contradict_2}\\
  &T^l_k=\frac{\left(1-\alpha_k\right)\omega_k L_k}{\tilde f_k}\leq T^{p}.\label{Contradict_1}
	\end{align}
\end{subequations} 
To proceed, we re-adjust the offloading policy and computation resource allocation. The offloading percentage and remaining portion are updated to $(1-\delta_1)\alpha_k$ and $(1+\alpha_2)(1-\alpha_k)$, respectively. The computation resource locally is changed to $(1-\delta_3)\tilde f_k$. After updating the offloading policy, the computing time cost at the server end is $\tilde T^p=(1-\delta_1) T^{p}$, which is less than $T^{p}$ when $\delta_1>0$. Now, we will prove that the updated strategy can meet other constraints.
After updating the offloading policy, the time cost $\tilde T^l_k=\frac{1+\delta_2}{1-\delta_3}T^l_k$ and energy consumption $\tilde E^l_{k} = (1-\delta_3)^2 (1+\delta_2)E^l_{k}$. Then, we can find three small positive values $\delta_1$, $\delta_2$, and $\delta_3$, which meet
 \begin{subequations}\label{contradiction}
	\begin{align}
	&\tilde T^l_k<T^{p},\label{Contradiction_1}\\
 &\tilde E^l_{k} \leq  (1-\delta_1)P_kT^p,\label{Contradiction_2}\\
 &(1-\delta_1)\alpha_{k} + (1+\delta_2)(1-\alpha_{k})= 1.\label{Contradiction_3}
	\end{align}
\end{subequations} 
This can be easily proved by algebraic simplification. In this paper, we omit this proof process. Inequality (\ref{Contradiction_1}) avoids that the computation time locally is larger than the previous maximum. Inequality (\ref{Contradiction_2}) ensures that the energy consumption can meet constraint (\ref{ob_g}). Inequality (\ref{Contradiction_3}) ensures that the sum of the offloading and locally computing percentages equals 1. As a result, we can search for a new transmit strategy to satisfy $\tilde T^p<T^p$ and all constraints in the problem (\ref{linear_p2}), which contradicts the initial assumption. We thus prove the first conclusion.

Next, we use a similar method to prove the second conclusion. Let $T^l_k$ and $T^l_{k'}$ denote the maximum and minimum among $\{T^l_1,\dots,T^l_K\}$. If we keep the other variables unchanged and curtail the offloading percentage of $k'$-th user. Let $(1+\delta_5)(1-\alpha_k)$ denote the local computation percentage at the user end. We can easily find a small positive value $\delta_5$ that can meet $(1+\delta_5)T^l_{k'}<T^l_k$ and other constraints. The data processing time cost is still $T^l_k$, but it can reduce data offloading and feedback time cost since the offloading data is reduced. This helps reduce the overall time cost, which contradicts the assumption. We thus prove the second conclusion.\hfill \QEDclosed

According to Claim 3, combining equations (\ref{Time_l}) and (\ref{Energy_l}), the energy consumption constraint in (\ref{ob_g}) can be recast as
\begin{equation}
\kappa \tilde f^3_k \leq P_k,~\forall k.
\end{equation} 
Meanwhile, the computation resource should not exceed its threshold $F_k$. Therefore, the optimal computation resource allocation with the closed-form expression at the user end is 
\begin{equation}
\tilde f^*_k = \min{\left(\sqrt[3]{\frac{P_k}{\kappa}}, F_k\right)},~\forall k.
\label{Closed}
\end{equation}

Similarly, as for the energy consumption at the BS end, we can observe 
\begin{align}
&\sum_{k=1}^{K}\frac{E^s_k}{T^p} \overset{(a)}= \sum_{k=1}^{K}\frac{\kappa \tilde f_kf^2_k\beta^2_k}{(1-\beta^2_k)}  \notag\\=& \sum_{k=1}^{K}\left(-\kappa \tilde f_kf^2_k +\frac{\kappa \tilde f_k}{2(1-\beta_k)}+\frac{\kappa \tilde f_k}{2(1+\beta_k)}\right).
\end{align}
where equation $(a)$ holds since $T^p=T^l_k$ for $\forall k$. Adopting the first-order Taylor expansion approximation, energy consumption constraint(\ref{ob_e}) can be recast to 
\begin{align}
 \sum_{k=1}^{K}\left(-\kappa \tilde f_kf^{(n)}_k\left(2f_k-f^{(n)}_k\right)+\frac{\kappa \tilde f_k}{2(1-\beta_k)}+\frac{\kappa \tilde f_k}{2(1+\beta_k)}\right)\leq P_b,
 \label{Taylor}
\end{align}
which can define a convex set since $\tilde f_k$ has been optimized with a closed-form expression. In inequality (\ref{Taylor}), superscript $(n)$ denotes iteration index. Similarly, constraint(\ref{ob_d2}) can be transformed to 
\begin{align}
\frac{\omega_k\left(1-\beta^{(n)}_k\left(2\beta_k-\beta^{(n)}_k\right)\right)L_k}{\tilde f_k}\leq T^p,~\forall k
\label{Time}
\end{align}

By using the above details, we can reformulate the problem (\ref{linear_p2}) as
\begin{subequations}\label{linear_p3}
	\begin{align}
&\min_{\mathcal{Q}_1,\mathcal{Q}_2}T^u+T^p+T^d,\label{ob_a2}\\
	\text{s.t.}\quad
	&\mbox{(\ref{ob_b}),~(\ref{ob_c}),~(\ref{ob_d}),~(\ref{ob_h}),~(\ref{Taylor}),~(\ref{Time}),~(\ref{ob_b2}),}\label{ob_b3}\\
        &\mbox{(\ref{ob_c2}),~(\ref{ob_e2}),~(\ref{ob_f2}),~(\ref{ob_g2}),~(\ref{ob_h2}),~(\ref{ob_i2}),}
	\end{align}
\end{subequations}
where $\mathcal {Q}_1$ and $\mathcal {Q}_2$ collect the inherent variables and auxiliary variables, respectively, \emph{i.e.}, $\mathcal {Q}_1=\{\mathbf{W},\mathbf{p},\mathbf{f},\mathbf{R},{\bm{\beta}},\mathbf{T}\}$ and $\mathcal {Q}_2=\{\mathbf{y}_{k,c},\mathbf{y}_{k,p},\mathbf{Y}_{k,m},\mathbf{\Phi}_{k,m}\}$. Note that variable $\tilde{\mathbf{f}}$ is removed since $\tilde f_k$ has been solved via (\ref{Closed}). All of the constraints have been transformed into convex sets except for constraints (\ref{ob_f2}), (\ref{ob_g2}), and (\ref{ob_h2}). In these three constraints, the non-trivial coupling between beamforming and auxiliary variables incurs non-convexity, making the optimization problem difficult to solve directly. However, we observe that the problem (\ref{linear_p3}) is convex when we fix the introduced auxiliary variable $\mathcal {Q}_2$. Meanwhile, when the beamforming matrices and vectors are given, we can derive the optimal auxiliary variable $\mathcal {Q}_2$ with closed-form expression via Claim 1 and Claim 2. Based on this fact, we treat the problem (\ref{linear_p3}) as a two-tier AO problem, where $\mathcal {Q}_1$ and $\mathcal {Q}_2$ are optimized alternately by the convex framework and closed-form expressions, respectively. At the $(n)$-th iterative, the optimal $\mathcal {Q}^{(n)}_2$ is given as follows.
 \begin{subequations}\label{close-form}
	\begin{align}
	 &\mathbf{y}^{(n)}_{k,c}=(I_{k,c})^{-1}\mathbf{h}^H_k\mathbf{p}^{(n)}_c,\\
  &\mathbf{y}^{(n)}_{k,p}=(I_{k,p})^{-1}\mathbf{h}^H_k\mathbf{p}^{(n)}_k,\\
&\mathbf{Y}^{(n)}_{k,m}=\left(\bm{\Omega}_{k,m}+\mathbf{H}^H_{k}\mathbf{W}^{(n)}_{k,m}\left(\mathbf{W}^{(n)}_{k,m}\right)^{H}\mathbf{H}_{k}\right)^{-1}\mathbf{H}^H_{k}\mathbf{W}^{(n)}_{k,m},\\
&\mathbf{\Phi}^{(n)}_{k,m}=\left(\mathbf{W}^{(n)}_{k,m}\right)^{H}\mathbf{H}_{k}\left(\bm{\Omega}^{(n)}_{k,m}\right)^{-1}\mathbf{H}^H_{k}\mathbf{W}^{(n)}_{k,m}.
	\end{align}
\end{subequations}

We summarize the proposed AO algorithm in Algorithm ~\ref{Alg.1}. Here are its crucial properties, including convergence, optimality, and complexity.
\begin{itemize}
\item  \emph{Convergence and Optimality}: Given an arbitrary feasible initial point $\mathcal{Q}_1$, our algorithm always seeks the globally optimal solutions in lines 3 and 4. This indicates that it can find the last feasible point after each iteration, so the objective value is non-increasing. Meanwhile, since the total time cost cannot be reduced infinitely, we thus deduce that our algorithm converges within finite iterations. The convergence indicates at least a local optimal solution.

\item \emph{Complexity}: In line 3, the computational burden mainly stems from $\mathbf{Y}^{(n)}_{k,m}$ and $\mathbf{\Phi}^{(n)}_{k,m}$ since the complexity solving other one-dimensional variables is almost negligible. For two matrices $\mathbf{W}_1\in\mathbb C^{ A_1\times A_2}$ and $\mathbf{W_2}\in\mathbb C^{ A_2\times A_3}$, the complexity of $\mathbf{W}_1\mathbf{W}_2$ is $\mathcal O\left(A_1A_2A_3\right)$. Thus, the complexity in line 4 is in order of $\mathcal{O}\left(A_{br}A_{ut}A_u\right)$. In the convex optimization process via CVX, the computational complexity is $\mathcal{O}\left(N^{3.5}\right)$, where $N$ is the number of variables. In line 4, our number of variables is $2KA_{ut}A_u+(K+1)A_{bt}+6K+3$. Thus, the per-iteration complexity is in order of $\mathcal{O}\left((KA_{ut}A_u+KA_{bt})^{3.5}\right)$. However, using the SDR-based algorithm, its complexity is in order of $\mathcal{O}\left((KA_{ut}^2+KA_{bt}^2)^{3.5}\right)$.
\end{itemize}

\begin{algorithm}[t]
	\caption{Alternating Optimization (AO) Algorithm} 
	\begin{algorithmic}[1]\label{Alg.1}
		\STATE Initialize $\mathcal{Q}^{(1)}_1$, $n=1$, and the maximum tolerance $\epsilon$.
		\WHILE { no convergence}
		\STATE \qquad Calculating optimal $\mathcal{Q}^{(n)}_2$ via (\ref{close-form});
		\STATE \qquad Solve problem (\ref{linear_p3}) via CVX and output optimal \\\qquad$\mathcal{Q}^{(n+1)}_1$;
  \STATE \qquad Update $n=n+1$;
		\ENDWHILE
		\STATE Output minimized total time cost $T^u+T^p+T^d$.
	\end{algorithmic}
\end{algorithm}

\section{Simulation results}\label{Section IV}
\begin{table}[t]
	\caption{Key simulation parameters}
	\begin{center}\label{Table I}
		\begin{tabular}{l l  p{13cm}}       
			\toprule  	\cmidrule(r){1-2}
			$\mathbf{Parameter}$ & $\mathbf{Value}$ \\
			\midrule
			\hline
			Cell radius & $100$~meters\\
			\hline
                Transmit antenna per user: $A_{ut}$ & $2$\\
			\hline
			Transmit antennas at BS: $A_{bt}$ & $4$\\
			\hline
			Receive antennas: $A_{br}$ & $2$\\
			\hline
			Number of users: $K$ & $8$\\
			\hline
			Transmit power per user: $P_{k}$ & $10$~dBm\\
			\hline
			Transmit power at BS: $P_b$ & $30$~dBm\\	
			\hline
			Computation capacity per user: $F_k$ & $3$~M cycles/s\\
			\hline
   			Computation capacity at BS: $F_b$ & $1$~G cycles/s\\
			\hline
			Path-loss at $d$~km from transmitter &$128.1+37.6\log_{10}\left(d\right)$ dB\\
			\hline
                Bandwidth & $10$~MHz\\
                \hline
			Background noise power & $-100$~ dBm\\
			\hline
			Cycles processing one bit task: $\omega_n$ & $297.2$~cycles/bit\\
                \hline
			Compression factor: $\epsilon$ & $0.5$\\
			\hline
			$\kappa$ & $10^{-24}$\\
			\hline
		\end{tabular}
	\end{center}
\end{table}

This section presents comprehensive simulations evaluating the proposed RS-based transmit approach and algorithm. The simulated healthcare network spans a circular area with the BS situated at the center, and users are randomly distributed within this region. The size of medical data varies randomly between $1$~Mbit to $5$~Mbit. Each simulation result is averaged over $100$ independent Rayleigh channel realizations. Unless specified otherwise, Table \ref{Table I} provides all the simulation parameters, primarily sourced from \cite{9309177,9184073,qiu2021computation}.

We benchmark our transmit scheme and algorithm (labeled as {\bf{RS-AO}}) against four baselines to assess its performance, summarized as follows comprehensively.
\begin{itemize}
\item {\bf{NOMA-AO}}: Each message is encoded into one stream in uplink data offloading and downlink result feedback. Each receiver with $(K-1)$-layer SIC removes all stronger interference before detecting its desired stream. 
\item {\bf{SDMA-AO}}: This benchmark uses the same encoding technique as the first. However, each receiver decodes its desired stream by treating other streams as noise.
\item {\bf{RS-SDR}}: This one utilizes uplink and downlink RS to offload and feedback medical data. However, the formulated optimization problem is done via the SDR-based iterative algorithm. This algorithm is from \cite{8392790}. Note that the SDR curve is the upper bound of its performance due to the rank constraints of semi-definite matrices being ignored during solving.

\item {\bf{RS-Cloud}}: This baseline offloads all of the collected medical data to the BS for further processing, \emph{i.e.}, the offloading percentage is equal to $100\%$.
\end{itemize}

\begin{figure}[tbp]
	\centering
	\subfigure[]{
		\begin{minipage}[t]{0.48\linewidth}
			\centering\includegraphics[width = 1.66in]{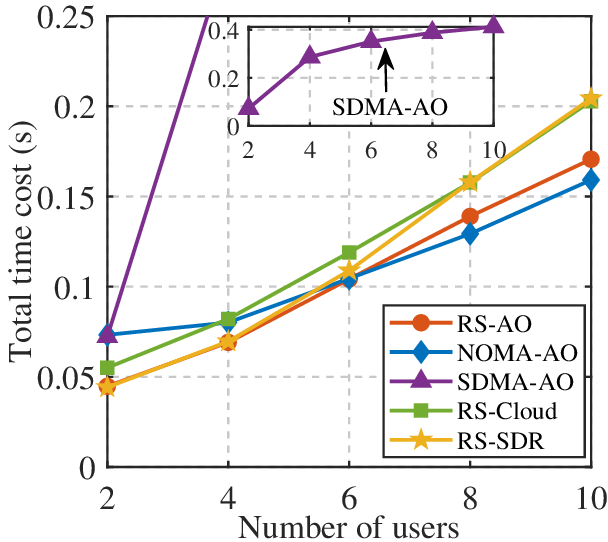}
		\end{minipage}
	}
	\subfigure[]{
		\begin{minipage}[t]{0.45\linewidth}
			\centering\includegraphics[width = 1.51in]{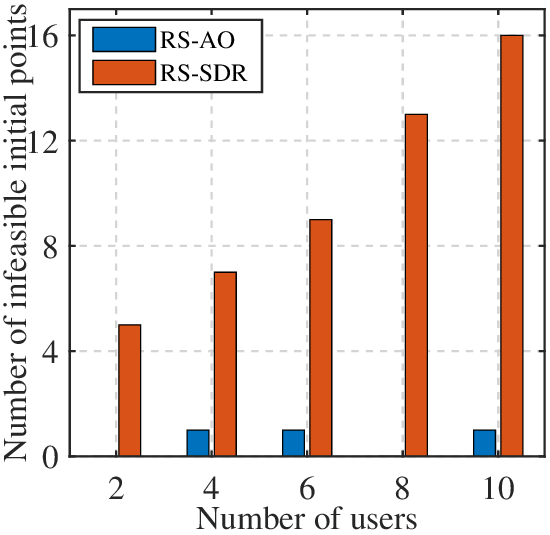}
		\end{minipage}
	}
	\centering
	\caption{(a) The total time cost versus the number of users. (b) The number of infeasible initial points versus the number of users.}
	\label{Fig_user}
\end{figure}

Fig.~\ref{Fig_user}(a) describes the total time cost versus the number of users, while Fig.~\ref{Fig_user}(b) records the number of infeasible initialization in $100$ channel realizations. The initial beamforming matrix and vector are randomly and independently generated, following Rayleigh distributions. As the number of users increases, so does the total time cost of all transmit schemes and algorithms. However, compared to SDMA, the time cost of RS and NOMA increases slowly. This case occurs since RS and NOMA can decode part or all stronger interference. Another interesting observation is that our RS scheme outperforms NOMA when $K\leq 6$ and exhibits the opposite phenomenon when $K\geq 8$. However, our RS approach performs closely with that of the NOMA. This phenomenon happens due to the following reasons. Using NOMA, the number of SIC layers increases linearly with increasing users. For example, when $K=6$ and $K=8$, each medical device must be equipped with $5$- and $7$-layer SIC, respectively. Its performance gains benefit from the increase in the number of SIC layers.

In contrast, the downlink RS always requires single-layer SIC. The performance of RS with $1$-layer SIC surpasses that of NOMA with $5$-layer SIC. This underscores the superiority of RS in managing co-channel interference. We observe that our algorithm produces two remarkable benefits compared to the classic SDR algorithm. Firstly,   our proposed algorithm achieves fuller exploration of the solution space than the SDR algorithm, and the performance gap gradually broadens (Fig.~\ref{Fig_user}(a)). Secondly, our algorithm adapts well to the random initial points, but this is not true for the SDR algorithm (Fig.~\ref{Fig_user}(b)). Also, the infeasible number of the SDR algorithm increases linearly. 

\begin{figure}[tbp]
	\centering
	\includegraphics[scale=0.6]{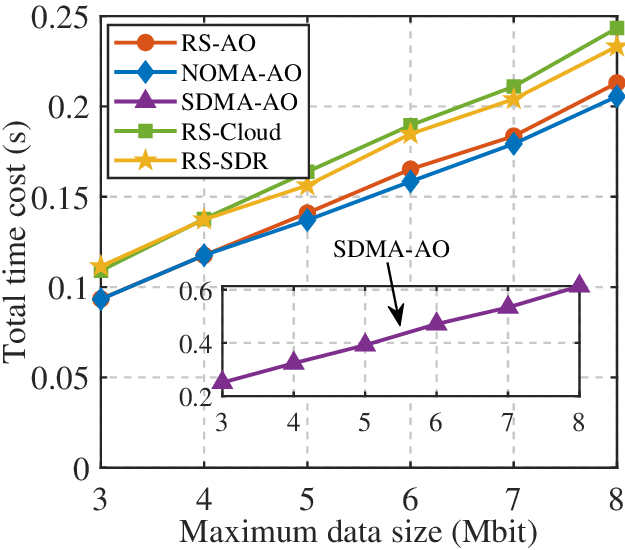}
	\caption{Total time cost versus maximum data size}
	\label{Size}
\end{figure}
Fig.~\ref{Size} presents the total time cost versus the maximum data size. This figure shows that fog computing helps alleviate healthcare networks' total time cost since it allows part of the collected medical data to be processed at the source node. The time cost gap between fog computing and cloud computing becomes clearer as data size increases. Compared to the SDR algorithm, our proposed algorithm achieves significant performance enhancements, about 13\%. These gains again highlight the effectiveness of the proposed algorithm. The time cost of our RS scheme is slightly higher than that of the NOMA when $K=8$. However, our transmit approach only needs a single-layer SIC for low-power medical devices. This caters to the evolving requirements of healthcare networks.

\begin{figure}[tbp]
	\centering
	\includegraphics[scale=0.55]{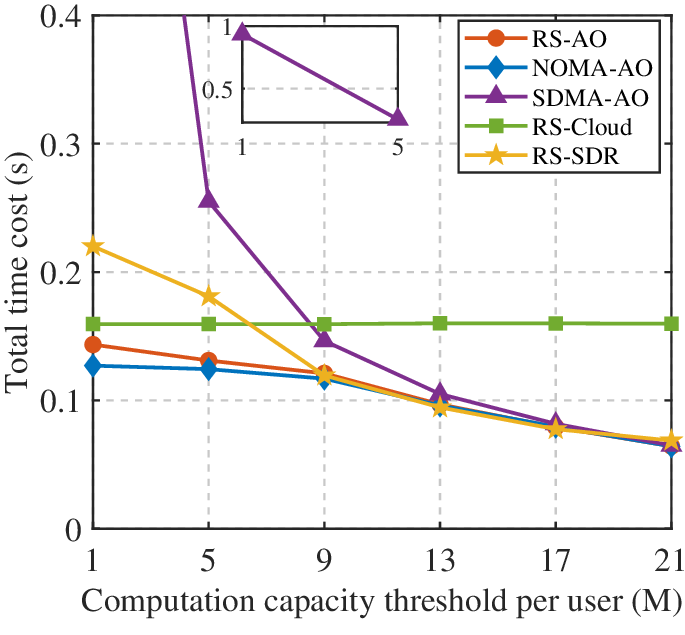}
	\caption{Total time cost versus computation capacity per user.}
	\label{capacity}
\end{figure}
Fig.~\ref{capacity} investigates the impact of the computation capacity threshold per user. As expected, the simulation result reveals that the time cost of cloud computing-enabled healthcare networks remains flat. This is because cloud computing-enabled networks offload all medical data to the BS for processing, so its performance is independent of the computation capacity per user. Besides cloud computing, other transmit approaches and algorithms degrade the time cost by enhancing user processing capability. Meanwhile, the performance gap between different transmit schemes and algorithms gradually disappears. The reason is that when the computation capacity is sufficient, users are more willing to process all the collected medical data than transmit them to the BS. 

\begin{figure}[tbp]
	\centering
	\includegraphics[scale=0.63]{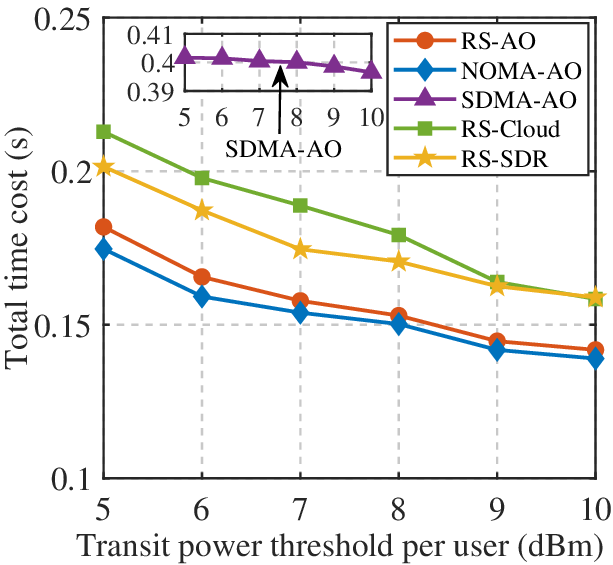}
	\caption{Total time cost versus transmit power.}
	\label{power}
\end{figure}

Fig.~\ref{power} simulates the total time cost versus the transmit power, where the transmit power at the BS increases from $20$~dBm to $30$~dBm. For example, an abscissa equal to $5$ means that $P_k=5$~dBm and $P_b=20$~dBm. The simulation result indicates that the performance of the SDMA is insensitive to the increase of the transmit power. This validates that SDMA yields the saturated rate when co-channel interference becomes excessive. However, our RS scheme alleviates the total time cost as the transmit power increases, overcoming the rate saturation phenomenon. This points to the flexibility of RS in managing inter-user interference.

\begin{figure}[tbp]
	\centering
	\includegraphics[scale=0.6]{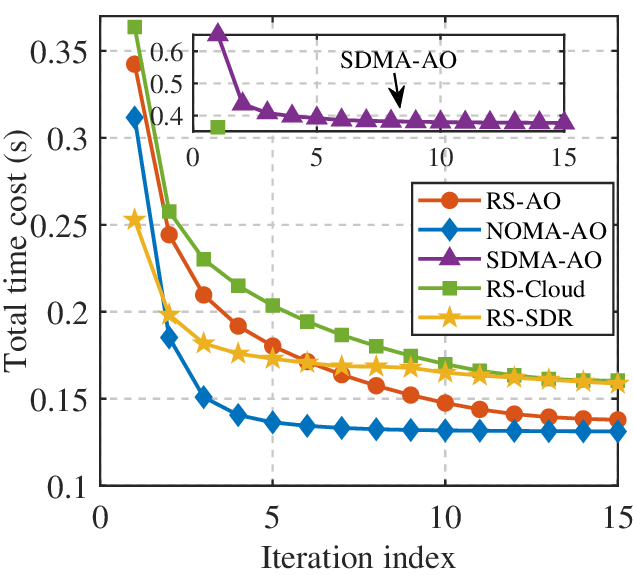}
	\caption{Total time cost versus iteration index.}
	\label{Convergence}
\end{figure}
Fig.~\ref{Convergence} shows the convergence behavior of the proposed algorithm. It converges to a steady value within a few iterations, validating the convergence analysis in Section \ref{Section III}. Our algorithm requires more iteration numbers than the SDR algorithm, which empowers it to explore the solution space better. Moreover, the computation complexity of our algorithm is much lower than that of the SDR algorithm. 

\section{Conclusion}\label{Section V}
We introduced a novel uplink and downlink RS transmit scheme tailored for fog computing-enabled IoMT systems. This scheme harnesses RS interference management capabilities to alleviate offloading challenges and feedback delays. We jointly optimized offloading policies, computation resource allocation, uplink and downlink beamforming, and common rate allocation to minimize total time costs.
Since the formulated problem is non-convex, we employed the quadratic transform method to generate accurate surrogates for the non-convex objective function. Subsequently, we utilized the first-order Taylor expansion to convert non-convex energy consumption and processing time into convex constraints. These transformations enabled us to address the reformulated problem as a two-tier AO problem.
Simulation results underscored the efficacy of our transmit scheme and algorithm, showcasing significant gains over several benchmark scenarios. 

Our study offers valuable insights for future research avenues. For instance,  RS-enabled task migration among multi-servers can be researched to balance server burdens and alleviate processing delays. Additionally, since user mobility may affect the accuracy of CSI estimation, leveraging the robustness of RS to address such scenarios and devising corresponding resource allocation algorithms would be worthwhile pursuits.

	\ifCLASSOPTIONcaptionsoff
	\newpage
	\fi
	
	\bibliographystyle{IEEEtran}
	\bibliography{references}
	
\end{document}